# On the Energy Dependence of Galactic Cosmic Ray Anisotropies in the Very Local Interstellar Medium

Romina Nikoukar [ORCID],[1] Matthew E. Hill [ORCID],[1] Lawrence Brown,[1] Stamatios M. Krimigis [ORCID],[1]
Robert B. Decker,[1] Konstantinos Dialynas [ORCID],[2] Jozsef Kota,[3] Edmond C. Roelof,[1]
Scott Lasley,[4] Douglas C. Hamilton,[4] V. Florinski,[5] J. Giacalone,[6] John Richardson,[7] and
Merav Opher[8, 9]

[1]*Johns Hopkins University Applied Physics Laboratory, 11100 Johns Hopkins Rd., Laurel, MD, USA*

[2]*Academy of Athens, Athens, Greece*

[3]*University of Arizona, Tucson, AZ, USA*

[4]*University of Maryland, College Park, MD, USA*

[5]*University of Alabama in Huntsville, Huntsville, AL, USA*

[6]*University of Arizona, Tucson, AZ, USA*

[7]*MIT, Cambridge, MA, USA*

[8]*Boston University, Boston, MA, USA*

[9]*Radcliffe Institute for Advanced Studies at Harvard University, Cambridge, MA, USA*

## ABSTRACT

We report on the energy dependence of galactic cosmic rays (GCRs) in the very local interstellar medium (VLISM) as measured by the Low Energy Charged Particle (LECP) instrument on the *Voyager 1* spacecraft (V1). The LECP instrument includes a dual-ended solid state detector particle telescope mechanically scanning through 360° across eight equally-spaced angular sectors. As reported previously, LECP measurements showed a dramatic increase in GCR intensities for all sectors of the $\geq$ 211 MeV count rate (CH31) at the V1 heliopause (HP) crossing in 2012, however, since then the count rate data have demonstrated systematic episodes of intensity decrease for particles around 90° pitch angle. To shed light on the energy dependence of these GCR anisotropies over a wide range of energies, we use V1 LECP count rate and pulse height analyzer (PHA) data from ≥ 211 MeV channel together with lower energy LECP channels. Our analysis shows that while GCR anisotropies are present over a wide range of energies, there is a decreasing trend in the amplitude of second-order anisotropy with increasing energy during anisotropy episodes. A stronger pitch-angle scattering at the higher velocities is argued as a potential cause for this energy dependence. A possible cause for this velocity dependence arising from weak rigidity dependence of the scattering mean free path and resulting velocity-dominated scattering rate is discussed. This interpretation is consistent with a recently reported lack of corresponding GCR electron anisotropies.

## 1. INTRODUCTION

Galactic Cosmic Rays (GCRs) are high-energy charged particles that originate far outside the heliosphere through various mechanisms such as particle accelerations at plasma shocks associated with supernovae or stellar wind collisions (e.g. Ackermann et al. 2013). Although they were discovered in 1912 (Hess 1912) , the GCR acceleration mechanisms, composition, and their filtration and modulation in the interstellar medium and heliosphere are not yet well understood (e.g. Hamaguchi et al. 2018).

*Voyager 1* (V1) is the first human-made object travelling outside the heliosphere into the very local interstellar medium (VLISM), and hence V1 charged particle measurements are critical to determine the source and dynamics of GCRs. Upon exiting the heliosphere on 25 August 2012, the Low Energy Charged Particle (LECP) and Cosmic Ray Subsystem (CRS) instruments on-board V1 observed orders of magnitude decreases in anomalous cosmic rays (ACR) along with significant increases in the GCR count rate (Krimigis et al. 2013; Stone et al. 2013). Since then, the time series of LECP count rates above ≥ 211 MeV have exhibited several episodes of reduced proton intensity and time-varying depletion of particles with pitch angle close to 90° (Krimigis et al. 2013; Hill et al. 2020). The pitch angle information from LECP data indicates a second-order anisotropy, with decreased intensity in the direction perpendicular to the average magnetic field (we delve further into this behavior herein). The anisotropic depletion of GCRs were also accompanied by plasma wave events, weak shocks, and variations in the magnitude of the magnetic field (Gurnett et al. 2015). Similarly, omni-directional (≥ 20 MeV) proton-dominated measurements from CRS show up to a 3.8 % intensity reduction (Rankin et al. 2019). By analyzing CRS Bi-directional (≥ 70 MeV) and unidirectional (~18 to ~70 MeV) proton-dominated measurements during various spacecraft orientations, including magnetometer roll calibrations and 70° offset maneuvers, they characterized this anisotropy as a "notch" in an otherwise uniform pitch-angle distribution of varying depth and width centered about 90° in pitch angle space. In another study, Rankin et al. (2020) compared GCR ions and electrons. They found very similar profiles for protons in the 30-70 MeV and in the ≥ 70 MeV energy range, indicating a rather weak energy dependence of these GCR events. Electrons, on the other hand, responded to the shocks in the same way as ions did, but did not show any noticeable depletion. The reason of this unexpected disparity between the anisotropies of ions and electrons is still an open question (which we address in the Discussion section).

To date, several physical interpretations and models have been proposed to explain these anisotropic features of GCR flux depletion. Roelof et al. (2013) suggest that the anisotropy is due to a trapped configuration between two magnetic field compression regions. Numerical simulations from Jokipii & Kota (2014); Kóta & Jokipii (2017) suggest that a gradual compression, followed by a slow weakening of a magnetic field, may account for both the episodic increases and the anisotropic reduction in GCR count rates. While Roelof et al. (2013) recognizes the role of the spatial variation of the magnetic field, Kóta & Jokipii (2017) emphasizes temporal evolution and the role of adiabatic cooling in the weakening field. A more comprehensive calculation by Zhang & Pogorelov (2020), simulating a spherical shock propagating through the heliopause into the VLISM, reached similar conclusions. This calculation considered only protons at one single energy (100 MeV), and did not address energy dependence of the GCR event. It is also worth mentioning that Strauss & Fichtner (2014) suggest "heliopause shadowing" as the source of the anisotropy using a model for the perpendicular diffusion coefficient peaking at the 90°. This model predicts a gradual decrease in anisotropy away from the heliopause. In view of unusual timing of these anisotropic variations and transient disturbances in the interstellar medium observed by magnetic field and plasma oscillations, in an observational study using measurements from ACE, New Horizon and *Voyager 1* and *2*, Hill et al. (2020) present an alternative escaping particle picture in which a disturbance interacting on only one side of an VLISM field line results in the observed time-dependent anisotropies.

In this work, we present an observational study using *Voyager 1* LECP measurements to investigate the anisotropy in GCR fluxes at different energy ranges. The energy dependence of the anisotropy of GCR fluxes can provide essential observational constraints on numerical simulations and theoretical prediction of GCR modulation, filtration, and transport in the VLISM. In Section 2, we provide a brief description of the LECP instrument and different types of measurements used in this study and the techniques we employ in order to quantify the anisotropy. In Section 3, we present our findings on anisotropy at different energy ranges. In Section 4, we provide a summary of this investigation and offer implications of our results in terms of future theoretical development on GCR modulation and filtration in the VLISM.

## 2. OBSERVATIONS

The LECP on *Voyager 1* employs a set of several solid state detectors, absorbers, and small magnets designed to measure differential intensities of ions from ~ 40 keV to ~ 60 MeV nuc$^{-1}$ and integral measurements above 211 MeV, and electrons of ~ 26 keV to >10 MeV (Krimigis et al. 2003, 2013; Decker et al. 2005; Dialynas et al. 2021). Pitch angle information is obtained through mechanical rotation of the detectors. The main detectors look within a single scan plane that is rotated 360 degrees, stopping at 8 different look sectors, labeled 1 to 8. The lower energy detectors have full width view cones of about 45 degrees. Sectors 1 and 5 are aligned approximately perpendicularly to the average magnetic field as measured by the magnetometer instrument (e.g. Burlaga et al. 2020), and hence the data from these sectors correspond to particles with ~ 90° pitch angle.

The LECP instrument consists of two subsystems, the Low Energy Particle Telescope (LEPT) and the Low Energy Magnetospheric Particle Analyzer (LEMPA). In this work, we utilize the data from the LEPT subsystem, as it is termed in the LECP instrument paper (Krimigis et al. 1977), but hereafter we restrict LEPT and HEPT to refer to the low and high energy ends of the particle telescope, respectively. The two types of measurements that we use in this study include count-rate data and pulse-height analyzer (PHA) data. Count-rate data at each channel are obtained through analog circuitry (pulse height discriminators and coincidence circuitry) when particles pass through or deposit energy at specific detectors. Detailed PHA "event data" are available for a subset of measured particles, restricted due to telemetry limitations. For this type of data, first the pulse height analysis is completed based on a priority scheme detailed in (Krimigis et al. 1977; Hill 1998), and then energy channel data number (DN) values corresponding to pulse height are determined by the analog-to-digital converter which is part of the PHA circuitry. The channel values are indicative of energy deposits in four of the detectors. Hence, while rate data are determined by on-board LECP analog electronics, PHA measurements are determined based on ground analysis software post-processing, and contain detailed information of the amount and location of energy deposited at each detector.

## 2.1. *Uni-directional Flux Correction*

The dual-headed nature of the LECP particle telescope introduces some measurement challenges that we need to address in our analysis. For example, the CH31 GCR channel ($\geq$ 211 MeV) on LECP is composed of coincidence measurements from two 2.5-mm thick solid state detector (D3 and D4) along with anticoincidence measurements from the eight-element, $360°$ anticoincidence shell. Any particle that deposits energy in these two detectors, but not the anticoincidence shell, hence triggering the analog circuitry, will increment the rate count for this channel. Because this telescope is dual-headed, with logic that ignores the signals from the detectors at the entrances of LEPT (D1 & D2) and HEPT (D5), it is not possible to determine the direction of incoming particles in a convectional way, because the desired foreground particles entering from either side produce nearly identical signals in D3 and D4 and less intense signals in the remaining detectors. Here, we describe the procedure that we use to untangle this effect and determine the direction of incident particles. Ignoring the effect of the inactive volume of the instrument (e.g., the housing, electronics, spacers, etc.) the geometry of the coincidence and anticoincidence detectors associated with CH31 results in two roughly cone-shaped fields of view, but D3 and D4 are not laterally positioned in the middle of the anticoincidence shell, rather they are offset closer to the HEPT end of the telescope. Because of this offset, the geometry factors for the opposing heads are significantly different ($\frac{G_{LO}}{G_{HI}} \neq 1$) (Roelof et al. 2013). We were able to calculate the geometry factors based on the LECP dimensions and confirm the resulting values using in-flight data during an isotropic period in the LISM. This analytical technique amounts to solving a non-linear set of equations as discussed below. Figures 1(a) and (b) show the observational geometry for the incident particles from opposite directions (where $J_A$ and $J_B$ represent a pair of anti-aligned aperture-averaged differential intensities in units of cm$^{-2}$sr$^{-1}$s$^{-1}$MeV$^{-1}$) for the LECP 8-sector configuration. **R** and **T** show two axes from the RTN coordinate system, in which R is the radius vector from the Sun to the spacecraft, T is in the direction of solar rotation and N completes a right-handed system. The measured, oppositely directed particle counting rates $R_A$ and $R_B$ (units of counts s$^{-1}$) from the LEPT and HEPT sides (when the motor steps into the opposite, 180$^{\circ}$ offset configuration) can be written in terms of the constituent intensities and geometric factors $G_{LO}$ and $G_{HI}$ (units of cm$^2$sr) as

$$R_A = J_A G_{HI} + J_B G_{LO} \tag{1}$$

$$R_B = J_B G_{HI} + J_A G_{LO} \tag{2}$$

Solving these equations for flux entering the sector **A** yields

$$J_A = \frac{G_{HI} R_A - G_{LO} R_B}{G_{HI}^2 - G_{LO}^2} \tag{3}$$

with a corresponding expression for sector B. Therefore, we are able to determine the net incoming flux at one direction only when the rates from both oppositely-oriented telescope positions are available. We refer to this net incoming flux as uni-directionally corrected flux (referred to as Uni-flux for the remainder of this article) once the ambiguity in direction is removed. The uncertainty on the Uni-flux for sector **A**, $\sigma_{J_A}$ can be computed as

$$\sigma_{J_A} = J_A \sqrt{\frac{1 + \beta^2}{2 \Delta T} \frac{R_A + R_B}{(R_A - \beta R_B)^2}} \tag{4}$$

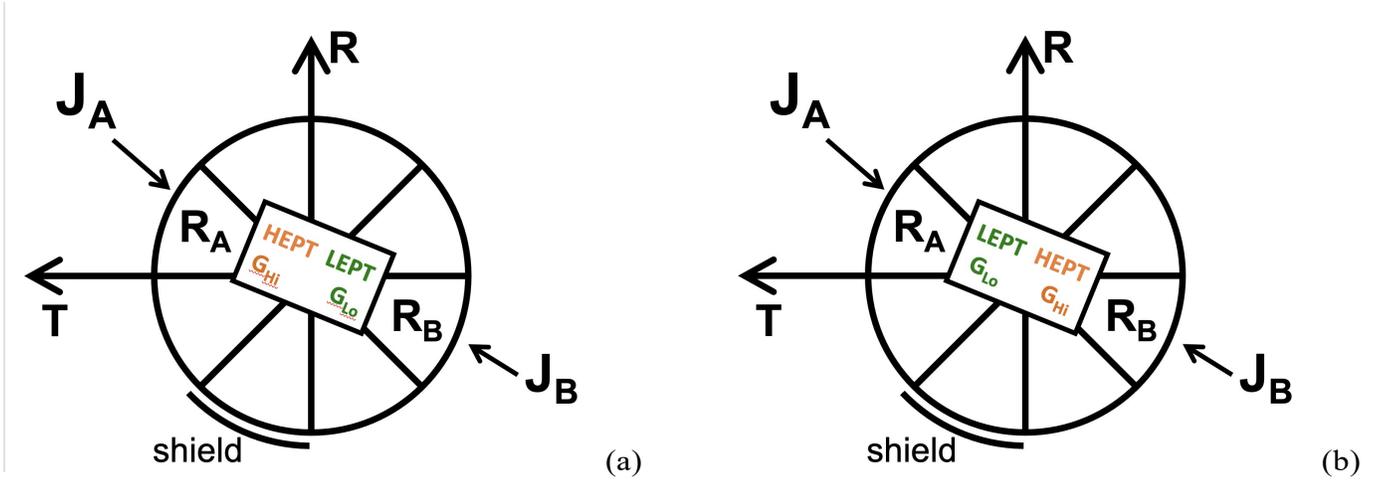

**Figure 1.** Two observational geometries for particles incident from opposite directions on the LECP particle telescope. Relating the measured rates to the fluxes, given the known, dissimilar LEPT and HEPT geometry factors, allows the uni-directional flux to be determined from each set of 180° out-of-phase measurements. The position of the sun shield is noted.

where $\beta = \frac{G_{Lo}}{G_{HI}}$ and $\Delta T$ notes the exposure time for LECP in the **A** configuration. We note that this process does assume that the time scale of the GCR intensity variations, especially the anisotropies, is much longer than the time scale of the 360° stepper motor movement. Since the former is from many days to 100s of days and the latter is approximately half an hour, this requirement is easily met.

### 2.2. *Anisotropy Determination*

The method we employ to determine the anisotropy is a Fourier component technique, where we fit angular distribution of LECP fluxes to a second-order Fourier expansion following Decker et al. (2005, 2015) as

$$j(\varphi) = A_0 + A_1 \cos(\varphi - \varphi_1) + A_2 \cos(2\varphi - 2\varphi_2) \qquad (5)$$

Note that the reference angle is taken from the **R** direction as shown in Figure 2(a).

The $A_0$ is a constant equal to the mean of measurements from all sectors. $A_1$ and $\varphi_1$ are the amplitude and phase corresponding to the first-order anisotropy, and $A_2$ and $\varphi_2$ are the amplitude and phase corresponding to the second-order anisotropy. Based on the harmonic fit, the first-order anisotropy is associated with the differences between the opposite sectors, while the second-order anisotropy is associated with the differences between a pair of opposite sectors with respect to the remaining sectors.

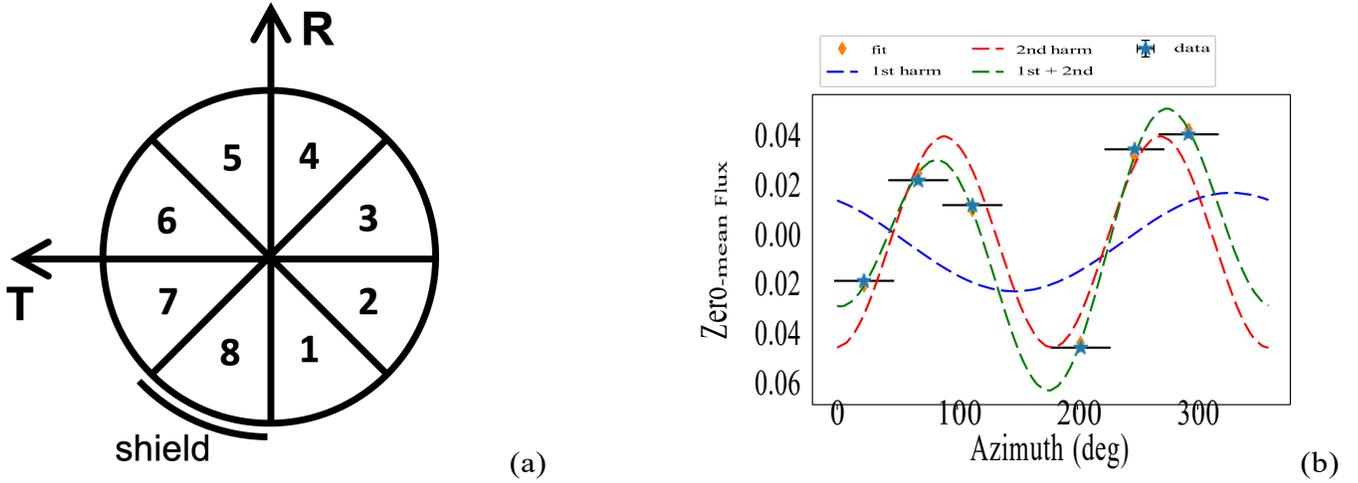

(a)                                                                              (b)

**Figure 2.** (a) LECP 8-sector configuration. The **R** axis defines the reference for the angular harmonics, taken in the usual counter-clockwise orientation. With this configuration, the center of the Sector 5 direction corresponds to a phase angle of 22.5°. (b) Example rate measurements of CH31 uni-directionally corrected fluxes (with mean subtracted) and the corresponding harmonic fit. The first and the second harmonic fits are shown in blue and red dashed curves, respectively. The final fit (sum of the harmonics) is shown in green. The harmonic fit yields 0.0192±0.004 and 327°±8.2° for the amplitude and phase of the first order anisotropy, and 0.0413±0.004 and 89° ± 2.6° for the amplitude and phase of the second order anisotropy, respectively.

The fit is conducted over the data from six sectors (1, 2, 3, 5, 6, 7). The data from sector 8, which holds the sun shield and its opposite sector (sector 4) are not included in the fit. The set of equations in (5) can be solved either analytically or numerically. Figure 2(b) shows a plot of example rate measurements from different sectors (uni-drectionally corrected fluxes) with zero mean for CH31 uni-directional fluxes. The data corresponds to a 9-day average centered at 2015-11-18. The harmonic fit yields 0.0192±0.004 and 327°±8.2° for the amplitude and phase of the first order anisotropy, and 0.0413±0.004 and 89° ± 2.6° for the amplitude and phase of the second order anisotropy, respectively. The observed azimuthal magnetic field angle is ∼ 286° ± 4° (Burlaga et al.

2018) for this time interval, roughly aligned with Sectors 3 and 7 (specifically $106°$ or $286°$ for anti-alignment and alignment, respectively,). Therefore, the 2nd order harmonic angle of $89° \pm 2.6°$ agrees sufficiently (within $17°$) with the interpretation that the depletion of flux (for particles from Sectors 1 and 5) is most significant at ~$90°$ pitch angles.

### 2.3. *Expansion to Lower Energy Channels*

The cosmic ray anisotropies at $90°$ pitch angles in the VLISM at V1, previously reported by Krimigis et al. (2013); Hill et al. (2020) using LECP measurements, apply to $\geq$ 211 MeV (CH31) count rate data. Our work here aims to explore the energy dependence of the GCR, thus we expand our analysis to lower energy LECP channels, i.e. 5 MeV - 20 MeV (CH16) and 21 MeV - 200 MeV (CH23). Unlike CH31, both CH16 and CH23 use a coincidence measure from the D5 detector near the HEPT entrance and anti-coincidence measurements from the detector beyond the stopping detector and hence do not suffer from the direction ambiguity that required the uni-directional flux correction.

Additionally, we use LECP PHA measurements to further divide each energy range into smaller intervals. For CH31, the incoming particles deposit their energy in the two thick detectors noted as D3 and D4. Figure 3(a) shows two smaller flux boxes that are being utilized for our analysis for CH31. The measured energy coordinates for these boxes are chosen such that their rates are nearly equal. The two smaller flux boxes for CH23 are overlaid on Figure 3(a). Protons with energies higher than 20 MeV deposit energy on the nominal track of CH23 (Box-0). As the energy of the incident particles increases, they deposit their energies on the return track (CH23 Box-1), and ultimately particles with energies higher than 211 MeV will deposit their energies on the designated CH31 area on D4 vs D3 detector plane. Note that the 40-200 MeV energy range covered in CH23 Box-1 is outside of the designed energy range and results from inefficiencies in the anti-coincidence veto. Figure 3(b) shows the smaller flux boxes for CH16 on D5 vs D4 detector plane. Table 1 provides a summary of V1 LECP channels, various PHA flux boxes, their corresponding energy range, and their nominal energy labels that we utilize in this study. We note that using the PHA raw data, which includes the detailed information on particle deposited energy and the associated hit position on the detectors, we can post process the data to collect all the box counts in each small flux box and compute the count rates associated with the PHA data, as follows.

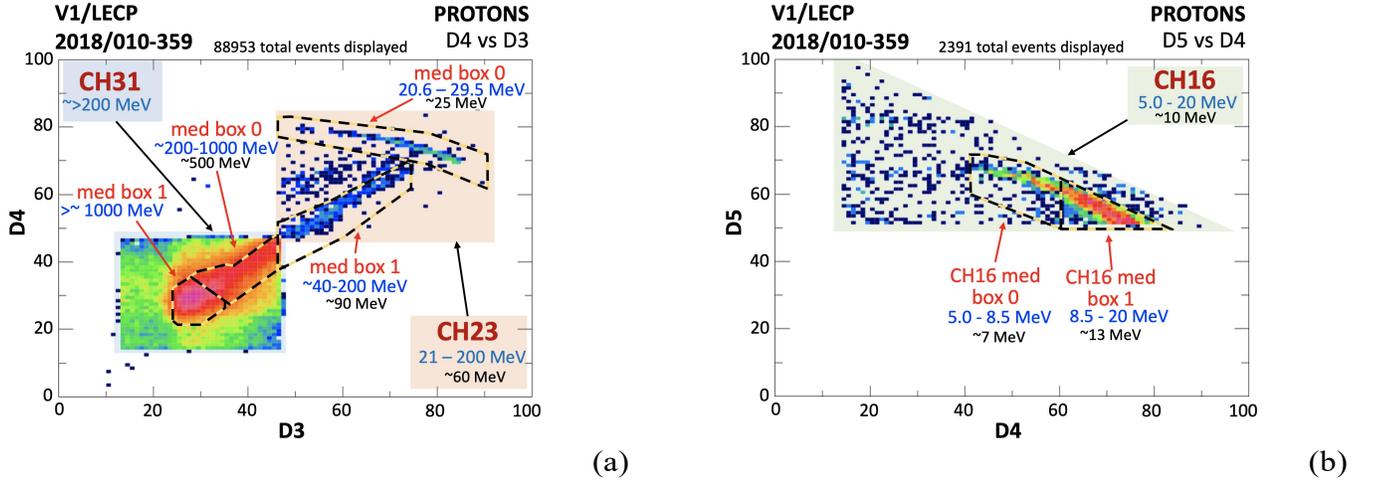

**Figure 3.** The larger, hardware-defined rate definitions and smaller ground-software based flux boxes are shown for the D4 vs D3 and D5 vs D4 PHA matrices, revealing the energy measurements for CH16, CH23, and CH31, where the ordinate and abscissa values are energy channel DNs. (a) Smaller flux boxes defined on D4 vs D3 for CH31 and CH23. (b) Smaller flux boxes for CH16.

**Table 1.** Definition of V1 LECP channels and flux boxes used herein.

| Channel | Flux Box[a] | Species | Energy Range | Energy Label[b] |
|---|---|---|---|---|
| CH16 | Big Box 16 | H | 5–20. MeV | 10 MeV |
| CH16 | Box 16-0 | H | 5.0–8.5 MeV | 7 MeV |
| CH16 | Box 16-1 | H | 8.5–20 MeV | 13 MeV |
| CH23 | Big Box 23 | H | 20.6–200 MeV | 60 MeV |
| CH23 | Box 23-0 | H | 20.6–29.5 MeV | 25 MeV |
| CH23 | Box 23-1 | H | 40–200 MeV | 90 MeV |
| CH31 | Big Box 31 | H[c] | ≥ 211 MeV | ≥ 211 MeV |
| CH31 | Box 31-0 | H[c] | 211–1000 MeV | 500 MeV |
| CH31 | Box 31-1 | H[c] | ≥ 211 MeV | ≥ 1000 MeV |

[a] Ground-software-based PHA analysis bin.

[b] Approximately geometric mean or lower bound.

[c] Minimal compositional discrimination, H expected to dominate.

$$r = \frac{X}{P} \frac{C}{\Delta T} \qquad (6)$$

$$\delta r = r \sqrt{\frac{1}{X} - \frac{1}{P} + \frac{1}{C}} \qquad (7)$$

where $X$, $P$, and $C$ represent the box counts, PHA counts, and Channel counts, respectively, and $r$ and $\delta r$ denote the count rate and the associated uncertainty for the PHA data. $\Delta T$ is the exposure time for acquiring data for the corresponding channel.

## 3. RESULTS

In this section, we present the anisotropy analysis results for LECP count rates and PHA measurements for $geq$ 211 MeV (CH31), 5 - 20 MeV (CH16) and 21-200 MeV (CH23) channels, showing the histories of ions with pitch angle close to 90° (Sectors 1/5) together with the time series of ions for the remaining pitch angles (Sectors 2/3/6/7), with "/" used as a delimiter in the list of sectors associated with each average.

Figure 4 shows the time series of averages (over 26-days) of Sectors 1/5 compared with averages from Sectors 2/3/6/7 for the LECP count rate data between 2012 and 2021. Figure 4a represents the extended time-series of $geq$ 211 MeV channel count rate data (box-car averages over 26 days) already reported in (Krimigis et al. 2013; Hill et al. 2020) showing a rapid increase in the count rates prior to the heliopause crossing in August 2012 as well as the anisotropic behavior of galactic cosmic ray in the VLISM. Several episodes of anisotropy, where significant decrease in sectors 1/5 count rates/fluxes are observed, are illustrated in these plots.

The time-series of the 26-day averages of Sectors 1/5 and 2/3/6/7 are shown in Figure 4(b) and (c). We note the significant decrease in count rates for these channels after the heliopause crossing of V1. This count rate reduction is more pronounced for 5 - 20 MeV (CH16) channel, where the count rates drop from ~ 2 counts/s to ~ 0.05 counts/s before and after heliopause crossing, respectively. The gradual decay of 3.4 - 17.6 MeV proton differential intensities at the heliopause was attributed to a flux tube interchange instability at the boundary (Krimigis et al. 2013), with GCRs slowly leaking out the flux tube into the heliosheath, thus decreasing their fluxes, whereas an outflow of 40-139 keV ions from the heliosheath (perpendicular to the magnetic field; S1 and S5) out to ~28 AU past the HP was recently shown in Dialynas et al. (2021). In Figure (4)d and e, we adjust the vertical axis range to properly visualize the count rate time variations for 20.6 - 200 MeV (CH23) and 5 - 20 MeV (CH16). These plots suggest that the relative time variation of 90° pitch angle particles compared to other pitch angle particles follows that of CH31 closely. The cyan lines in these plots indicate the average of non-perpendicular pitch angles after heliopause crossing to emphasize the depletion of perpendicular pitch angle particles. The uncertainty levels of measurements are also noted. Due to the decreased count rates, the relative error bars are higher for lower energy channels measurements compared to ≥ 211 MeV channel data.

The ratios of the data from Sector 1/5 over those of Sector 2/3/6/7 for the three channels are shown in Figure (5). The time series of (5 - 20 MeV) CH16 ratio shows the largest variability compared to those of other channels. This higher variability is due to the significantly lower count rates for CH16. At the heart of anisotropy episodes, CH23 seems to exhibit the smallest count rate ratio. Albeit these ratios provide a direct indication of the anisotropic behavior for the 90° pitch angle particles, they do not provide angular information. Figure (5)b and c show the normalized second-order anisotropy amplitude ($A2/A0$) and phase ($\varphi_2$), respectively. The normalized anisotropy amplitude is smallest for the CH31 rate data. The respective parameter is typically highest for CH23 while the associated variability is largest for CH16. The second-order anisotropy angle is centered close to 90° during the identified anisotropy episodes :2013-01 to 2013-07, 2014-01 to 2014-08, 2015-01 to 2016-12, and 2018-02 to 2020-01. During the isotropic periods, the second-order angular anisotropy can reach closer to 50° or 150° especially for CH16 measurements.

Next, we present the anisotropy results for the PHA data, big flux box measurements. The count rate for the big flux boxes is computed from the sum of small flux boxes parameters: box counts ($X_i$), PHA counts ($P_i$), channel counts ($C_i$) and exposure time ($\Delta T_i$), where $i$ indexes each flux box, as

$$r = \frac{\sum_{i=1}^{2} X_i \sum_{i=1}^{2} C_i}{\sum_{i=1}^{2} P_i \sum_{i=1}^{2} T_i} \qquad (8)$$

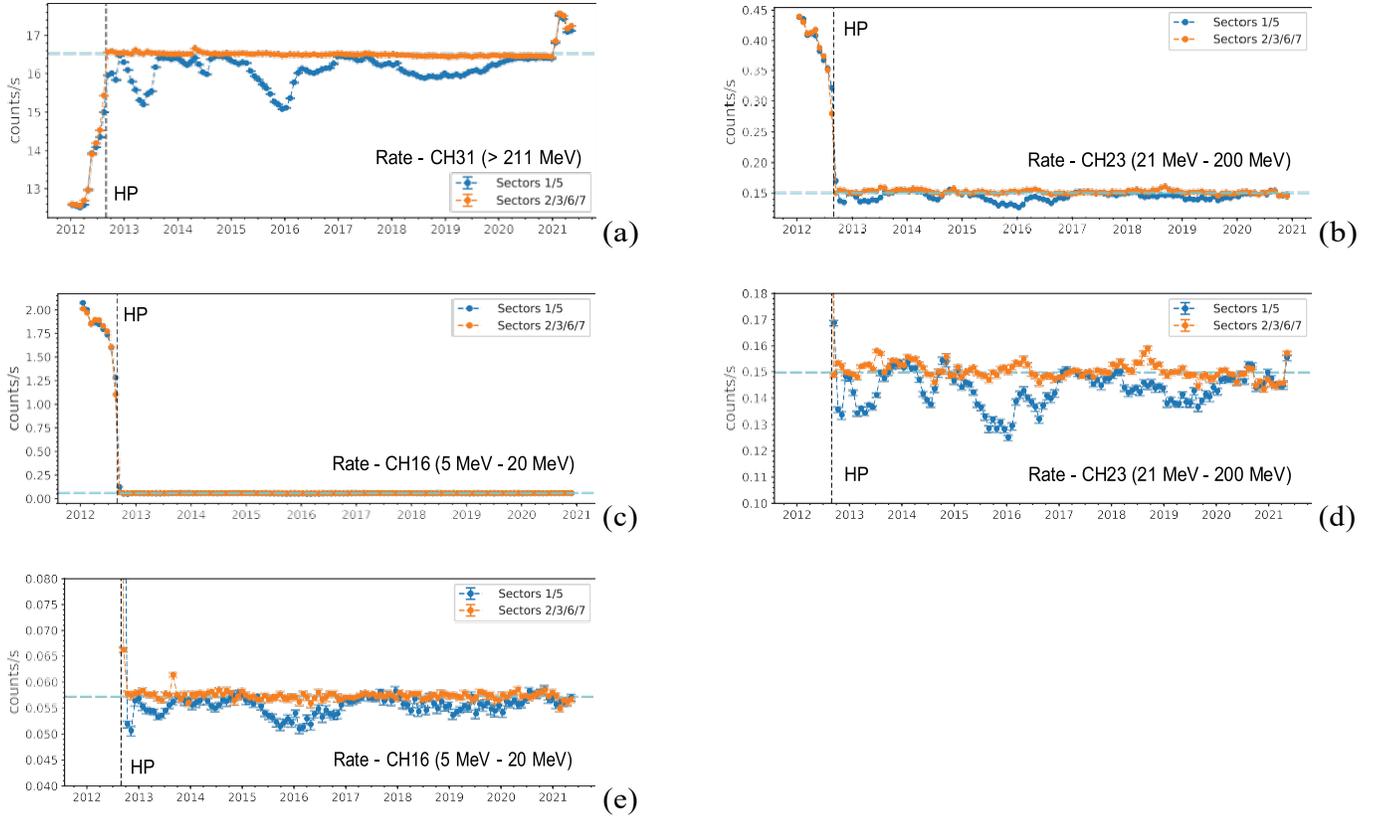

**Figure 4.** Rate data - Time series of rate measurements for LECP galactic cosmic ray channel (CH31) (a), and lower energy channels, CH23 (b) and CH16 (c). The zoomed in version of (b) and (c) is noted in (d) and (e), respectively, to emphasize the time profile after the heliopause crossing. In each plot, averages of sectors 1/5 (perpendicular to the average magnetic field) and averages of sectors 2/3/6/7 are shown with blue and orange curves, respectively. Error bars are also indicated.

with the small boxes shown in Figure 3(a) and (b) for CH31, 23, and 16. Figure 6 shows 26-day box-car averages of measurements from Sectors 1/5 and Sectors 2/3/6/7 for the big flux measurements of ≥ 211 MeV(CH31), 20.6 - 200 MeV (CH23), and 5 - 20 MeV (CH16) count rates. While the general trend of the time series matches that of the rate measurements, the measured uncertainties (error-bars) are significantly higher compared to the rate data. As explained earlier (Figure 4), the range on vertical axis in Figure 6(b) and (c) is adjusted for the proper visualization of the data after heliopause crossing. The ratio of measurements from Sector 1/5 over Sector 2/3/6/7 PHA big flux box data is shown in Figure 7(a), whereas the corresponding normalized amplitude and angle of the second anisotropy are shown in Figure 7(b) and (c). Similar to the rate data (Figure (5)), 20.6 - 200

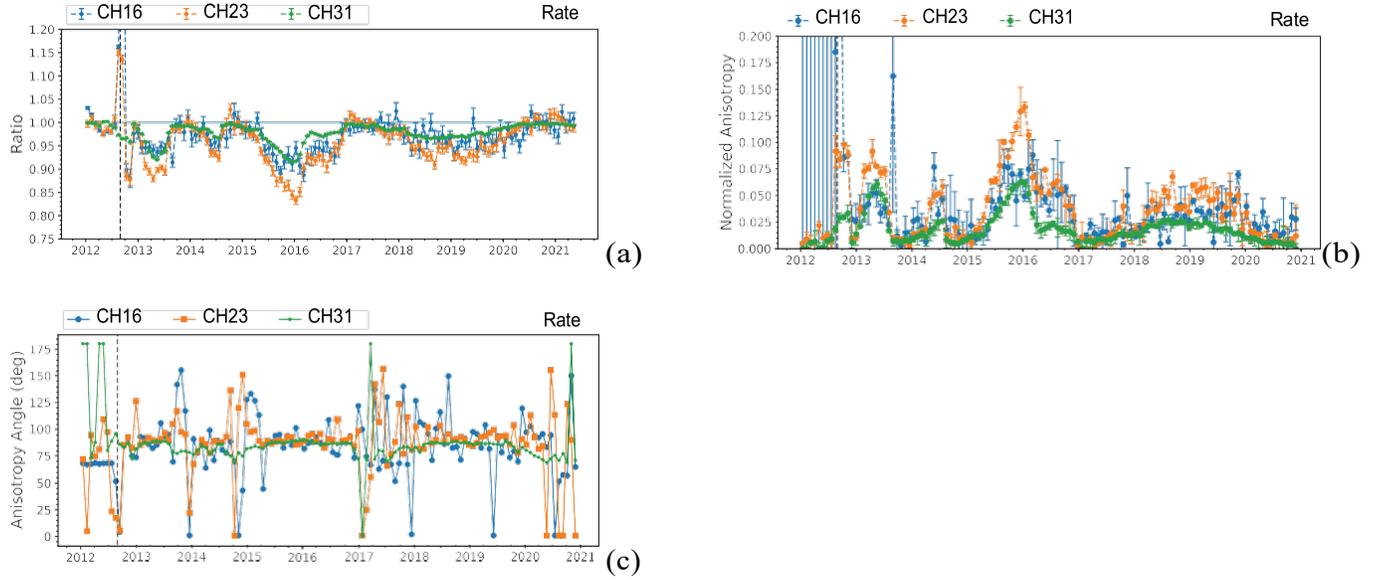

**Figure 5.** Rate data - Time series of the ratios of data from Sectors 1/5 over those of Sectors 2/3/6/7 for three different energies are shown in (a). The black vertical dashed and horizontal cyan line represent the *Voyager 1* crossing and ratio of 1, respectively. The normalized second-order anisotropy amplitude and phase are shown in (b) and (c), respectively.

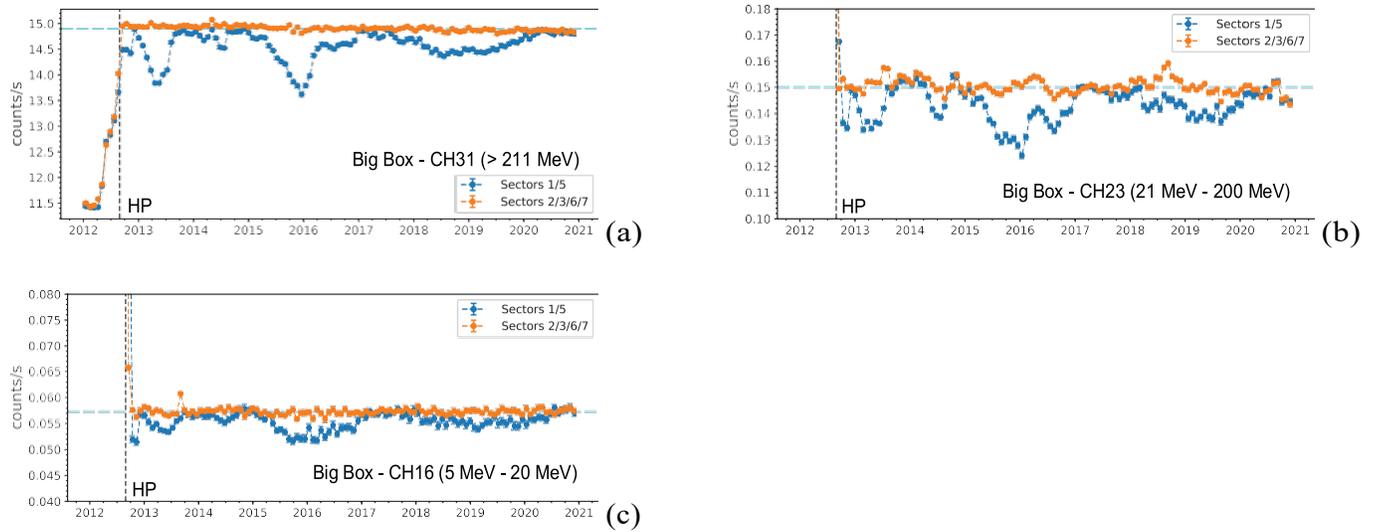

**Figure 6.** Big flux box data - Time series of Big flux measurements for LECP galactic cosmic ray channel (CH31) (a), and lower energy channels, CH23 (b) and CH16 (c). Similar to Figure (4), averages of Sectors 1/5 (perpendicular to the average magnetic field) and averages of Sectors 2/3/6/7 are shown with blue and orange curves, respectively.

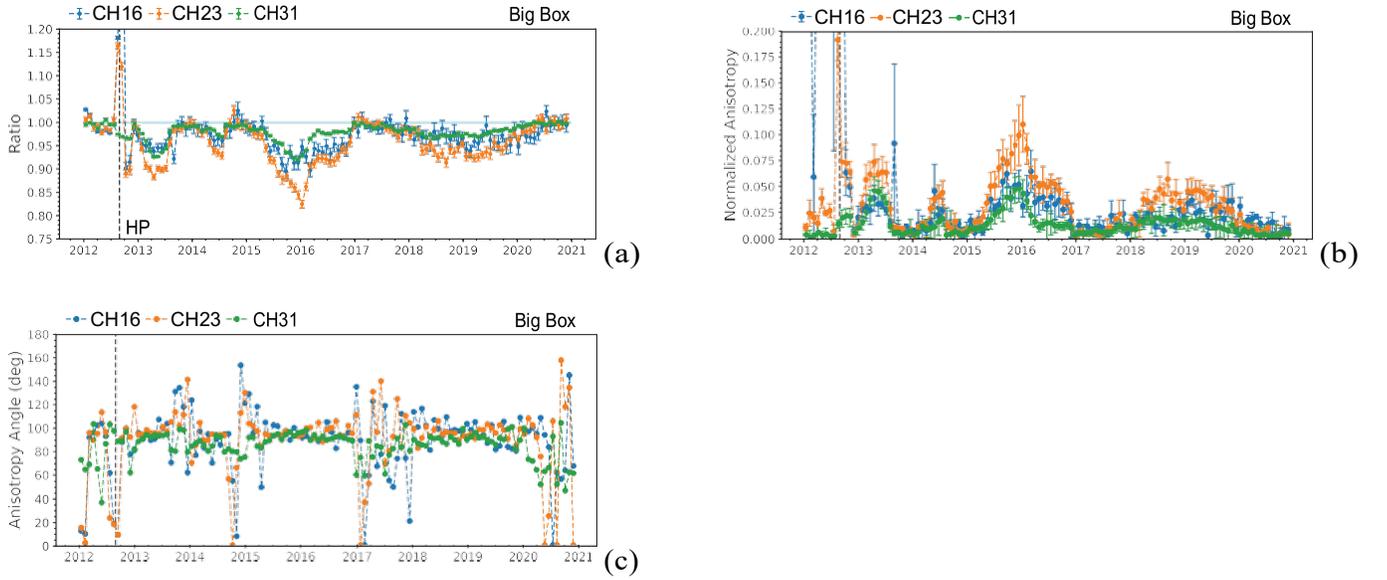

**Figure 7.** Big flux box data - Time series of the ratios of sectors 1/5 over sectors 2/3/6/7 for three different energies. The format and layout is similar to those of Figure (5).

MeV (CH23) big flux box data shows higher degree of anisotropy over short periods at the heart of anisotropy episodes. The data from the ≥ 211 MeV (CH31) channel exhibits the least anisotropic behavior. As a minor distinction between the two figures, we note higher uncertainty levels of the estimated PHA count rate ratios as well as second order anisotropy parameters. This is due to the aforementioned telemetry limit on PHA event data compared to the rate data, resulting in fewer counts in comparable PHA event-based quantities.

Finally, we examine the anisotropy of the galactic cosmic rays for the small PHA flux box count rates (Figure 3). In order to ensure credible statistics with the relatively limited PHA events, we compute averages of the data for longer time periods. The breakdown to smaller flux boxes allows for investigation of the time series trends of the 90° pitch angle particles with even finer energy resolution. An example is given in Figure 8, where the time-series of the data for different energy channels small flux boxes, 0 (left column) and 1 (right column), averaged over 200 days, are shown. Considering Figure 8 a and b, we find that the count rate depletion of Sector 1/5 associated with ~500 MeV (CH31 Box-0) is more significant than that of ~ 1000 MeV (CH31 Box-1). Similar to the rate and Big flux box data, we note the significant count rate reduction in the measurements from the two flux boxes of 20.6 - 200 MeV (CH23) (Figure 9 (c) and (d)), and an additional reduction in the two flux boxes of 5 - 20 MeV (CH16) (Figure 9 (e) and (f)). Moreover, we observe a slight reduction of count rates for the data corresponding to the lower energy flux Box-0 compared with the higher energy flux Box-1 for all three energy channels. Figure 8 (e) and (f) show lower count rates for Sectors 1/5 compared to Sectors 2/3/6/7, especially during the period around mid 2014 to 2017, although the timing of the reductions do not quite match with each other (We discuss a possible explanation for the misalignment below associated with Figure 10). Figure 9 illustrates the ratios of the measurements from Sectors 1/5 over those from Sectors 2/3/6/7 for the six small flux boxes.

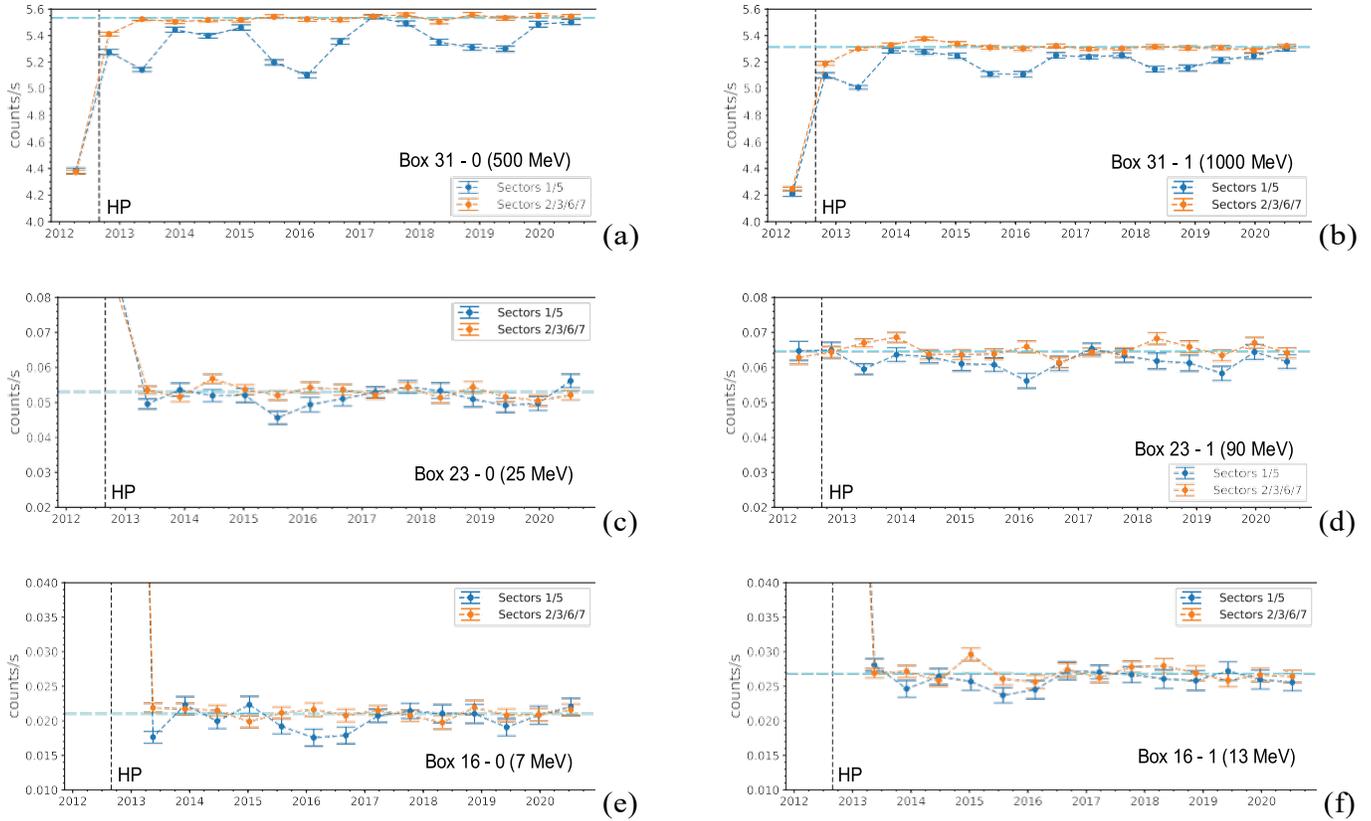

**Figure 8.** Small flux box - Time series of the ratios of data from Sectors 1/5 over Sectors 2/3/6/7 for small flux boxes for CH31 in the first row, CH23 in the second row, and CH16 in the third row. The data from flux boxes associated with lower energies (Box-0) in each channel are noted on the left column, whereas the data from the Box-1 for the three energy channels are shown in the right column. The cyan dashed line shows the averages of Sectors 2/3/6/7 after the 2012 heliopause crossing for each channel.

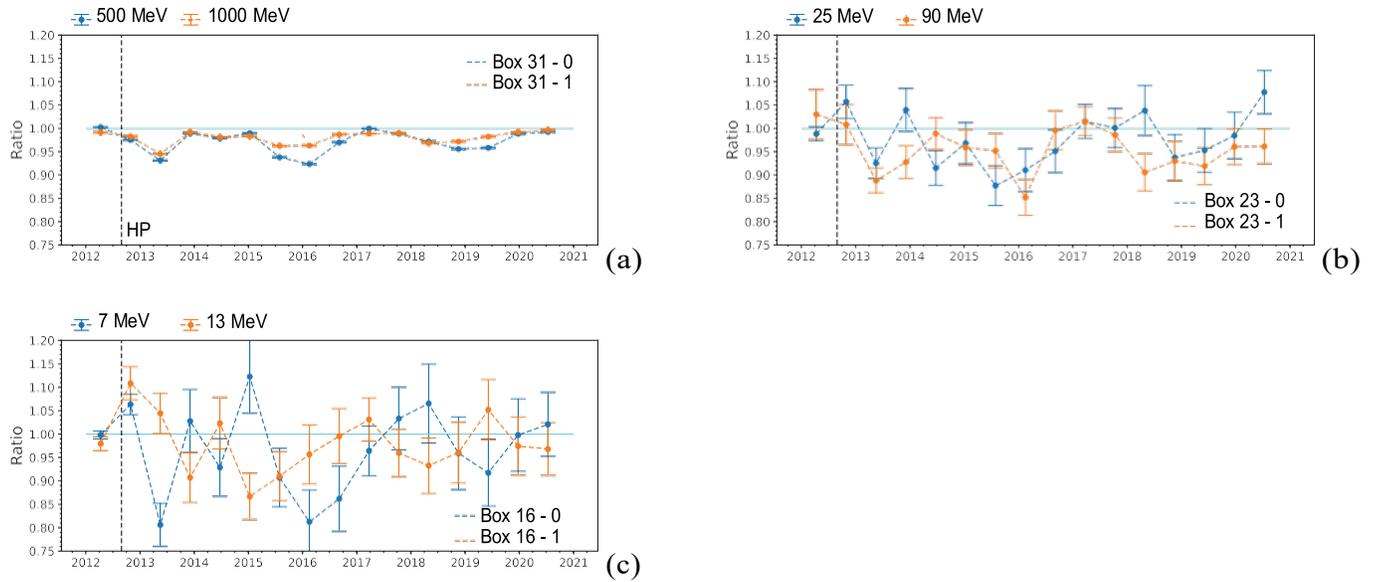

**Figure 9.** Time series of the intensity ratio of Sectors 1/5 over Sectors 2/3/6/7 for CH31 (a), CH23 (b), and CH16(c). In each plot, the ratio of data from flux box-0 and box-1 is noted in blue and orange, respectively. As expected the error bars increase for lower energy channels. CH16 Box-0 data shows a minimum ratio of $0.81 \pm 0.07$ during the anisotropy episode of 2016.

Similar to the plots from Figures 5 and 7, the ratios from the two CH31 flux boxes are typically higher compared to those of other channels. Breaking down to small flux boxes reveals an approximately 4% increase in the ratios for $\geq 1000$ MeV protons (CH31-Box1) compared to the lower energy portion $\sim$ 500 MeV (CH31-Box0). However, a major difference between these figures is that here the data from CH16 small flux boxes show lowest ratios reaching down to 0.8 for two anisotropy episodes (2013 and 2016 episodes). One reason for this discrepancy might be due to the misalignment between the time-series of the data for the two flux boxes of CH16. As described in Section 2, the rate and big flux box measurements are obtained via either the triggering of analog circuitry or combining the incident particles on specific detectors. Therefore, any misalignment in the time series for small flux boxes would reduce the magnitude of both the rate and big flux box measurements, and as such the corresponding overall intensity ratios are smaller compared to those of small flux box measurements.

In the final part of this section, we employ V1 Small flux box measurements to present the energy dependence of the galactic cosmic rays for three different anisotropy episodes. Following the methodology mentioned above, we first sum all the box counts, channel counts, PHA counts, and exposure time, and compute the count rates based on (8). We then compute the uni-directionally corrected fluxes for the CH31 flux boxes, and apply the fit to determine the second-order anisotropy parameters.

Figure 10 presents the second order normalized amplitude (left column) and angular (right column) anisotropy for three anisotropy episodes, i.e. 2013-01 to 2013-08, 2015-2016, and 2018-04 to 2019-06. Panels (a) and (b) illustrate the estimated anisotropy parameters for the episode between 2013-01 and 2013-08, which is about 212-day long. With the exception of CH16-Box1 (~ 13 MeV), the normalized amplitudes carry a decreasing trend with increasing energy, and the angular phase of the anisotropy is ~$95° \pm 10°$ (equivalently $275° \pm 10°$ since the second harmonic is periodic over $180°$). This result is in reasonable agreement (within error bars) with the azimuthal magnetic field of $< \lambda > = 292.5° + 1.4°$ (Burlaga & Ness 2016). The CH16 box-1 data does not follow the trend and seems to be more isotropic during this episode with an estimated second-order anisotropy angle of $45° \pm 43°$, and a normalized amplitude of $0.018 \pm 0.022$. One possible explanation for this result is the fact that this flux box is the only one in the current analysis that has an energy threshold that cuts through the track (as shown in Figure 3(b)). The vertical (D5) energy spread in the track is caused in part by the varying angles of incidence of particles entering the HEPT end of the telescope, where larger energy deposits correspond to larger angle relative to normal (bore-site-aligned) incidence. The trajectories of larger incident angles typically result in longer paths through the detectors, with proportionately more interactions between the incident particle and the silicon crystal, which, in turn, result in a larger energy deposit. This means that the particles most indicative of a given angular direction are preferentially excluded by the interfering energy threshold, thus complicating direct comparisons between CH16 box-1 and any other boxes.

The episode of 2015 to 2016 is the longest anisotropy episode. While the anisotropy phase is relatively close to $90°$ for all energies, confirming the depletion of the signals for the perpendicular pitch angle particles at all energies, the normalized amplitude decreases with increasing energy (panels c and d). The measurements from the two flux boxes associated with CH31 show the smallest normalized second-order amplitudes with $0.029 \pm 0.01$ and $0.017 \pm 0.006$ for ~ 500 MeV (CH31-Box0) and $\geq$ 1000 MeV (CH31-box1), respectively. Panels (e) and (f) of Figure 10, show the estimated parameters for the anisotropy period between 2018-04 to 2019-06. For this time interval, too, the normalized amplitudes are smallest for the CH31 flux boxes, while the amplitudes for the lower energy channels are approximately the same with relatively higher error bars. With the exception of Ch16-box1, all estimated phases of the 2nd-order anisotropy at this third period are lower (an average of $88°$) compared to the other two time intervals. This agrees well with the declining trend in the azimuthal magnetic field angle given in (Burlaga et al. 2020).

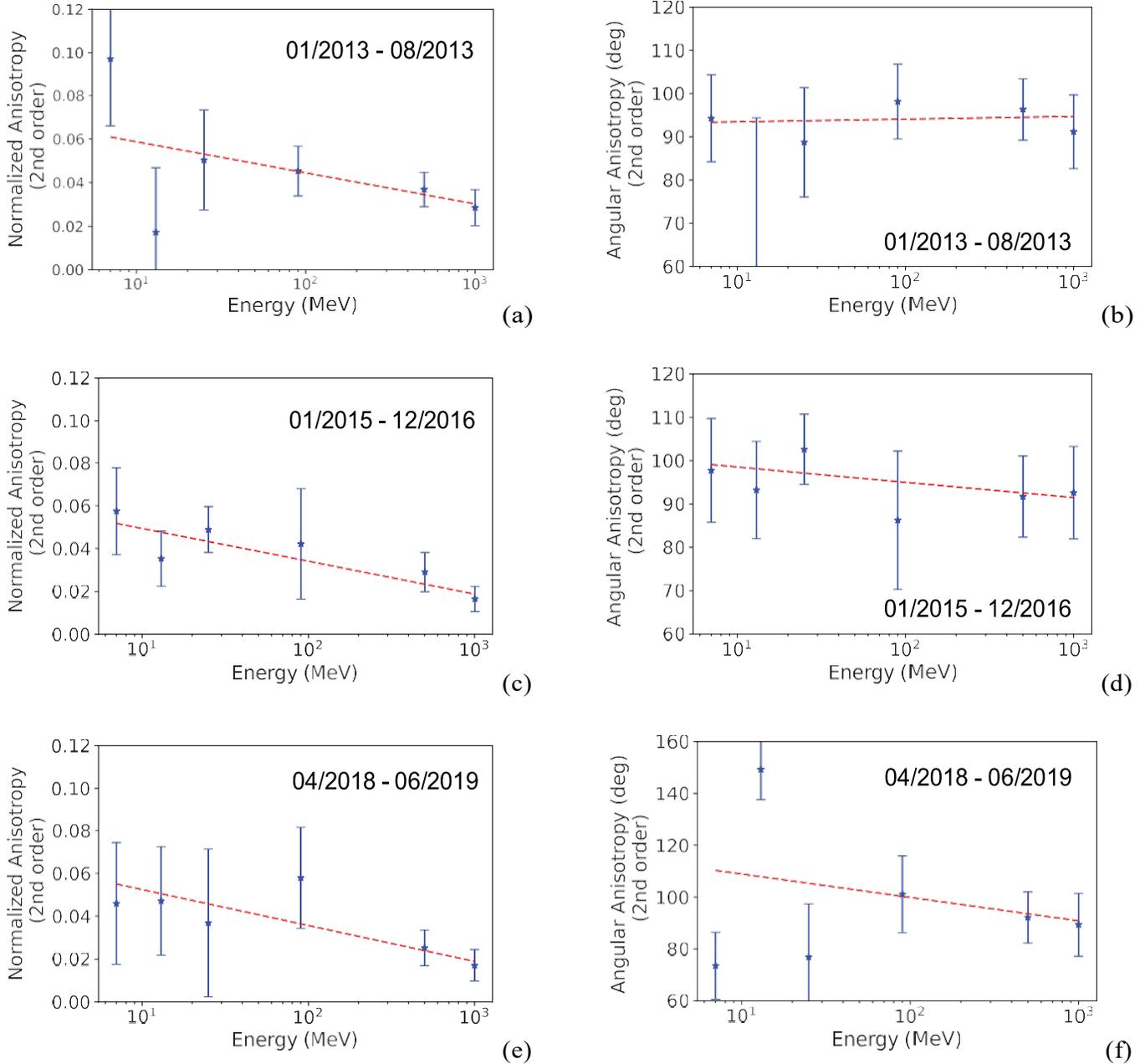

**Figure 10.** Second order anisotropy parameters as a function of energy during three anisotropy episodes: 2013-01 to 2013-08, 2015-2016, 2018-04 to 2019-06. The left column (a, c, e) shows the normalized amplitude, whereas the right column (b, d, f) provides the angular information for the second-order anisotropy. The dashed red line for each plot shows a linear fit to the data. The linear fits for the normalized amplitude as a function of the logarithm of energy have slopes -0.0143, -0.0153, -0.0168 in panels a, c, e, respectively. The negative slopes indicate a decreasing trend for the normalized amplitude with increasing energy for the three anisotropy episodes. The linear fits for the 2nd order anisotropy angle show near horizontal lines limited between 90° and 110° for all three anisotropy episodes confirming the depletion of particles perpendicular to magnetic field.

A linear fit to the normalized amplitude have slopes $-0.0143 \pm 0.0077$, $-0.0153 \pm 0.0038$, $-0.0168 \pm 0.0047$ in panels a, c, e, respectively. The negative slopes indicate a decreasing trend for the normalized amplitude with increasing energy for all three anisotropy episodes.

## 4. DISCUSSION

We provided a detailed observational investigation of the galactic cosmic ray anisotropies in the VLISM as a function of energy, using the *Voyager 1* LECP observations. Both the rate data and Big flux box PHA measurements show a higher degree of the second order anisotropy for the LECP lower energy channels ($\leq$ 60 MeV) compared to the LECP ($\geq$ 211 MeV) (CH31) uni-directional corrected measurements. The re-binning of the LECP PHA boxes to small flux box data confirms a stronger anisotropic behavior for low energy ions (~ $10 - 20$ MeV). For the LECP integral channel CH31 ($\geq$ 211 MeV), the re-binning of PHA shows a higher degree of anisotropy for energies ~500 MeV compared to $\geq$1000 MeV, for ions perpendicular to the magnetic field. For PHA analysis, a longer integration time of the data was required to the accumulate sufficient counts for the anisotropy study. Our analysis showed a decreasing trend in the amplitude of second-order anisotropy with increasing energy during anisotropy episodes with a slope of ~ -0.015.

Numerical simulations (Jokipii & Kóta 2014; Kóta & Jokipii 2017, e.g.) provide evidence that these anisotropies might be related to the heliosphere transient events indicating that the a gradual compression, followed by the a slow weakening of the magnetic field may account for the pitch angle GCR anisotropic decreases. According to the model, particles near $90^0$ pitch angles become trapped and cool down in the expanding field. The longer they are trapped the more energy they lose resulting in reduced intensity, where the spectrum is falling. The magnetic field observed at V1 has decreasing trends during the depletion events (see Gurnett et al. (2015)), which suggest that the depletion of GCRs is connected with the weakening of the field. For brevity, the model uses several simplifying assumptions, some of which may not hold for the long-lasting depletion events. The magnitude of the depletion is expected to depend on the spectral exponent of GCR, steeper spectrum should result is larger depletion.

The Kóta & Jokipii (2017) model considers purely magnetosonic variations in a simplified field-geometry, which could also lead to some modifications in the resulting GCR depletion. More importantly the model assumes scatter-free motion of GCRs along the magnetic field lines so that the adiabatic invariant $p_\perp^2/B$ remains conserved. This assumption might lead to the inability of the model to predict any energy dependence for the resulting anisotropy. We note that the more complex simulation of Zhang & Pogorelov (2020) includes a weak scattering, too. This work, however, considered only protons at one single energy (100 MeV), and did not address the energy dependence. The primary result of our analysis is the identification and characterization of the energy dependence of the anisotropy for V1 GCR ions during the anisotropy episodes in the VLISM. It is notable that although this energy dependence is pronounced in protons, the relationship that this has to charge and mass is not at all clear since Rankin et al. (2019) found that 3-105 MeV electrons do not exhibit any evidence of episodic anisotropy variations. We consider the situation in which it is the electron mass alone, not charge, that results in the lack of electron anisotropies, resulting in a requirement (for a monotonically varying anisotropy) that small mass and high energy both lead to minimal anisotropies. This rules out gyroradius as the controlling quantity, in addition to ruling out kinetic energy, charge, or mass, each taken alone. But a predominantly velocity-dependent anisotropy would still be consistent with the observations. Considering that the rate $\nu$ of pitch angle scattering (for isotropic scattering) scales as velocity $v$ over the mean free path $\lambda$, (where $\lambda$ depends in general on rigidity), we do obtain a scattering rate dominated by velocity if the rigidity dependence of the mean free path is weak. In the limit of a rigidity-independent mean free path the scattering rate becomes proportional to velocity with the mean free path setting the constant of proportionality, thus $v = v/\lambda$. The physical interpretation, put simply, is that for a constant mean free path the faster particles experience more frequent scattering than the slower particles and this serves to increasingly obscure the anisotropy as velocity increases, which means that the very fast electrons would show little anisotropy and the much slower protons would show more anisotropy, but with faster (and higher energy) protons showing a decreasing anisotropy as a function of energy, as we observe. A more advance modeling study to show the effect of pitch-angle scattering on energy dependence of the anisotropies is the subject of future work.

## 5. ACKNOWLEDGMENTS

This work was supported by the Voyager Interstellar Mission, NASA grant NNN06AA01C, and the NASA DRIVE science center, Our Heliospheric Shield, NASA grant 18-DRIVE18_2-0029, 80NSSC20K0603. M.O. was supported as well as the Fellowship Program, Radcliffe Institute for Advanced Study at Harvard University.